\documentclass[twoside,12pt]{article}
\usepackage{epsfig}
\usepackage{graphics}

\newcommand{\be}{\begin{equation}}
\newcommand{\ee}{\end{equation}}
\newcommand{\bea}{\begin{eqnarray}}
\newcommand{\eea}{\end{eqnarray}}

\begin{document}

\title{ Determination of the neutrino mass \\ by electron capture 
in 163 Holmium and \\ the role of the three-hole states in 163 Dysprosium.
} 
\author{Amand Faessler$^{1}$, Christian Enss$^{2}$, Loredana Gastaldo$^{2}$,
F. \v{S}imkovic$^3$, \\
\\
$^1$Institute of Theoretical Physics, University of Tuebingen, Germany\\
$^2$Kirchhoff Institut f\"ur
 Physik, University of Heidelberg, Germany\\
$^3$JINR, 141980 Dubna, Moscow Region, Russia and \\
Comenius University, Physics Dept., \ $SK-842 15$ \  Bratislava, 
Slovakia.}
\maketitle
\begin{abstract} 
$^{163}Ho$ to $^{163}Dy$ is probably due to the small Q value 
of about 2.5 keV the best case to determine the neutrino mass by electron capture. 
The energy of the Q value is distributed between the excitation of Dysprosium 
(and the neglected small recoil of Holmium) and the relativistic 
energy of the emitted neutrino including the rest mass. 
The reduction of the upper end of the deexcitation spectrum 
of Dysprosium  below the Q value allows to determine the neutrino mass. 
The excitation of Dysprosium can be calculated in the sudden approximation of the 
overlap of the electron wave functions of Holmium minus the captured electron and 
one- , two-, three- and multiple hole-excitations in Dysprosium. 
Robertson [R. G. H. Robertson, Phys. Rev. C91, 035504 (2015) and arViv: 1411.2906] and Faessler and Simkovic [Amand Faessler, Fedor Simkovic, accepted for Phys. Rev. C, March 2015 and arXiv: 1501.04338] have calculated the influence of the
two-hole states on the Dysprosium spectrum. 
Here for the first time  the influence of the three-hole states on the 
deexcitation bolometer spectrum 
of 163 Dysprosium is presented. The electron wave functions and the 
overlaps are calculated selfconsistently in a fully relativistic and 
antisymmetrized Dirac-Hartree-Fock approach in Holmium and in Dysprosium.
The electron orbitals in Dy are determined including the one-hole states 
in the selfconsistent iteration. The influence of the three-hole states 
on the Dy deexcitation (by X-rays and Auger electrons) can hardly be seen. 
The three-hole states seem not to be relevant for the determination of the electron neutrino mass.

\end{abstract}

\vspace{1cm}
\section{Introduction}

One of the most pressing problems in particle physics is the determination of 
the absolute value of the neutrino mass. The single beta decay, specifically 
the Triton decay, can give the electron antineutrino mass \cite{Drexlin} 
(present \ upper\ limit \cite{Otten}:  $m_{\bar{\nu}} \ \le \ 2.2 \ eV)$, 
while the neutrinoless double beta decay yields the mass of the 
effective Majorana electron neutrino mass \cite{Fae2}, if the transition 
matrix element can be calculated reliably. At the moment the upper limit 
for the Majorana neutrino mass from the double beta is about 0.3 eV.

The electron neutrino mass can be determined by electron capture. 
The most favorable case due to the small Q value of about 2.5 keV seems to be capture in $^{163}Ho$ (Z=67)  to 
atomic excited states in $^{163}Dy$  (Z=66). 
The energy of this Q value is divided into the excitation of the Dysprosium atom, 
 (plus the recoil of the Holmium atom, which can be neglected,) and into the 
energy carried away by the electron neutrino (kinetic energy and rest mass). 
The deexcitation spectra (X rays and Auger electrons) of all excited states 
in the Dysprosium atom have the upper end  
at the Q value minus the neutrino rest mass.
The determination of the electron neutrino mass from 
electron capture is more complicated than the determination 
of the antineutrino mass from the Triton decay, since more parameters have to be fitted to the data at the upper end of the spectrum. Under the 
assumption, that one excited Dy resonance with a Lorentzian line shape 
determines the behavior at the upper end,  
one has to fit simultaneously four parameters to the experiment after the detector response is folded into the theoretical spectrum: (1) The neutrino mass, (2) the energy distance of the leading resonance to the Q value, (3) the width  and (4) the strength of the resonance. Originally only one-hole excitations in the Dysprosium 
atom have been included in the theoretical determination of the spectrum \cite{de,Fae3}.
Recently Robertson \cite{Robertson} and with major improvements Faessler and Simkovic  \cite{Fae4} 
have included two-hole excitations in Dysprosium. The main improvement of ref. \cite{Fae4} over ref. \cite{Robertson} was, 
that the electron wave functions for Holmium and for Dysprosium have been calculated 
in these nuclei (Z=67 and Z=66) selfconsistently in a fully relativistic and antisymmetrized 
Dirac-Hartree-Fock approach \cite{Grant, Desclaux, Aukudinov} and not in Xenon (Z=54) as 
by Carlson and Nestor \cite{Carlson1, Carlson2}, results 
used by Robertson \cite{Robertson}. Due to the additional 
occupied states in Holmium and Dysprosium more 
two-hole states are allowed. In addition the electron   
orbitals in Dysprosium are determined selfconsistently 
with the appropriate  hole \cite{Fae4}. The finite size 
of the nuclear charge distribution is included using the Fermi 
parametrization determined by electron-nucleus scattering. 
The theoretical formulation was derived in second quantization 
including automatically the antisymmetrization and the overlap and exchange corrections.  

In the present work this formulation is extended for the first time 
to the three-hole states in Dysprosium including again all the improvements 
of the description of the two-hole states of ref. \cite{Fae4}. 

\vspace{1cm}

\vspace{1cm}

\section{Description of Electron Capture and \\ the Excitation of the Three-Hole States.}

The calorimetric spectrum of the deexcitation of $^{163}Dy$ 
after electron capture in $^{163}Ho$ can be expressed according to refs. \cite{de} and \cite{Fae3,Fae4}: 
 
\begin{eqnarray}
 \frac{d\Gamma}{dE_c} \propto \sum_{i = 1,...N_\nu}(Q - E_c)
\cdot U_{e,i}^2\cdot\sqrt{(Q-E_c)^2 -m_{\nu,i}^2} \\
*\sum_{h'=b',b'(p^{-1'}q),b'(p_{1'}^{-1}q_{1'})(p_{2'}^{-1}q_{2'})}
 \lambda_{0}B_{h'} \frac{\Gamma_{h'}}{2\pi} 
\frac{1}{(E_c - E_{h'})^2 +\Gamma_{h'}^2/4} \label{decay}
\end{eqnarray}                                     

The second sum over the excited states h' in Dy includes the one-hole states b', the two-hole states $b'(p^{-1'}q)$ and the three-hole states 
$b'(p_{1'}^{-1}q_{1'})(p_{2'}^{-1}q_{2'})$. 
With the Q value $Q \  = \ 2.3 \ to \ 2.8 \ keV $ \cite{Blaum, 
Anderson, Gatti, Ra, Audi} and a  recommended  
value \cite{Wang}  $Q = (2.55 \pm 0.016)$  keV.  The ECHo collaboration 
\cite{Ra} measured for the Q value 
for electron capture in $^{163}Ho$ to $^{163}Dy$:

\be
Q(ECHo) = 2.80 \pm 0.08\  keV.
\label{QECHo}
\ee
 
$U_{e,i}^2$ is 
the probability for the admixture of different 
neutrino mass 
eigenstates $i\ =\ 1,..N_\nu$  into the electron neutrino  
and $E_c$ is the excitation energy of final Dysprosium.
$B_h$ with the same orbital quantum numbers b = b' in Ho and Dy 
are the overlap and exchange corrections \cite{Fae3}; $\lambda_{0}$ contains 
the nuclear matrix element squared \cite{bam}; 
$ E_{h'}$ are the one-, two- and three-hole excitation energies in Dysprosium. 
$ \Gamma_{h'}$ is their widths \cite{Fae3}.

We assume, that the total atomic wave function 
can be described by a single Slater determinant. 
$ B_b $ takes into account the overlap and the exchange 
terms between the parent $ |G> $ and the daughter atom in 
the state $ |A'_{b'}> $ with a hole in the electron 
state $|b'>$. We use the sudden approximation as Faessler et al. \cite{Fae1}. 
$ B_b$ 
for the electron capture probability from the state b relative to the capture from $3s_{1/2}$ 
with one hole in b' in the Dysprosium atom is given in eq. (\ref{Bff}) 
in the Vatai approximation \cite{Vatai1, Vatai2}. But the numerical value 
used here are calculated with the full overlap and 
exchange corrections of Faessler et al. \cite{Fae3}. 

\begin{eqnarray}
 B_b = \frac{|\psi_{b}(R) <A'_{b'}|a_b|G>|^2}{|\psi_{3s1/2}(R)|^2} 
 = P_b \cdot \frac{|\psi_{b}(R)|^2}{|\psi_{3s1/2}(R)|^2}  
\label{Bff}
\end{eqnarray}

For two-hole final states one has to multiply eq.(\ref{Bff})   
with the probability to form a second hole characterized by the quantum numbers "p". 
One has to replace $<A'_{b'}|a_i|G>$ by \\ $<A'_{b',p'; q'}|a_i|G>$
with the two electron holes b' and p' and the additional electron particle q' in Dysprosium
 above the Fermi surface F. These expressions are derived in reference \cite{Fae4}. The formulas 
for three-hole states can be obtained with the help of Wick's theorem \cite{Wick} from: 

\begin{eqnarray}
  P_{p'_1,p'_2/b'} = |<A'_{p_1',p_2',b'}|a_b|G>|^2 = \hspace{4.5cm} \nonumber \\ 
	\sum_{q_1',q_2' > F}|<0| a'_{q_2'}a'_{q_1'}a'_{Z'=Z-1} ...a'_{p_2'+1}a'_{p_2'-1}...
	a'_{p_1'+1}a'_{p_1'-1}...a'_{b'+1}a'_{b'-1}...a'_1  a_b 
	a_1^{\dagger} a_2^{\dagger} a_3^{\dagger}... a_Z^{\dagger} |0>|^2 \label{Wick2}
\end{eqnarray} 

With the Holmium ground state:

\be
  |G> =  a_1^{\dagger} a_2^{\dagger} a_3^{\dagger}... a_Z^{\dagger} |0> \label{G}
\ee

The probability for forming three hole states can be calculated in the sudden approximation 
with the help of Wick's theorem \cite{Wick}.

To evaluate the probability for two electron 
particle-hole states   
one sums incoherently over all unoccupied states $q_1'$ and $q_2'$ .  

\bea
P_{b', p_1, p_2} =  \sum_{q_1' < q_2' > F} |<p_{1 < F,Ho}|q'_{1 > F,Dy}>
<q'_{1 >F.,Dy}|p_{1 < F ,Ho}>\cdot \nonumber \\<p_{2 <F, Ho}|q'_{2>F, Dy}>
<q'_{2>F, Dy}|p_{2<F, Ho}>|
\cdot  \prod_{k=k'<F_{Dy} \neq b,p_1, p_2}|<k'_{Dy}|k_{Ho}>|^2 
\label{F22}
\eea

Here as stressed above the sum over $q_1'$ and $q_2'$ runs over the unoccupied bound and continuum 
states in Dy. The dependence on the quantum numbers b = b' originates from the fact, that the electron 
orbitals in Dysprosium are calculated selfconsistently  with an empty state b'. 
One can now use the completeness relation to shift the sum 
over $q_i'$ from the continuum to states, 
which one can calculate easier. One divides the completeness
relation into two pieces: up to the last occupied state below the Fermi Surface  F  
and all states above the last occupied state  
including also the continuum. 

\begin{eqnarray}
 1 = \sum_{q_i' < F} <p_i|q_i'><q_i'|p_i>  + \sum_{q_i'>F} <p_i|q_i'><q_i'|p_i>  \nonumber \\
\mbox{with:} \hspace{0.5 cm} i = 1\ \ and \ \ 2  \hspace{4cm} 
\label{comp}
\end{eqnarray}

The sum in eq. (\ref{F22}) is the last part of the completeness relation 
 (\ref{comp}) and one can transcribe (\ref{F22}) into:

\begin{eqnarray}
P_{p_1,p_2/b} =  \left( 1 - \sum_{q_1'<F} <p_{1,Ho}|q'_{1,Dy}>
<q'_{1,Dy}|p_{1,Ho}>\right) \cdot    \nonumber   \\
\left( 1 - \sum_{q_2'<F} <p_{2,Ho}|q'_{2,Dy}>
<q'_{2,Dy}|p_{2,Ho}>\right) \cdot 
\prod_{k=k' < F_{Dy}; \ne p_1,p_2, b} 
<k'_{Dy}|k_{Ho}>  
\label{P3}
\end{eqnarray}

An interchange of two electrons from the states $b',\  p_1'\  and\  p_2'$  yields due to 
the antisymmetrization a "-" sign. But for the probability 
the minus sign is irrelevant. 
In the literature one uses often the Vatai 
approximation \cite{Vatai1, Vatai2}: Exchange corrections 
have been neglected already in eq. (\ref{P3}). In addition one assumes, 
that the overlaps of electron wave functions in the parent and the daughter atom 
with the same quantum numbers can be approximated by unity. Typically the overlaps 
have values \cite{Fae3} 
$ <k'_{Dy}|k_{Ho}> \approx 0.999$ and $0.999^{65} \ \approx \ 0.94$. In the 
Vatai approximation \cite{Vatai2} one replaces this value by 1.0. The probability for two one-electron particle and one-electron hole states is the product. 
The probability for a second hole in 
$p_i'$ with a first hole in b' is:

\bea
P_{p_i/b'} = \left( 1 - \sum_{q_i'<F} <p_{i,Ho}|q'_{i,Dy}><q'_{i,Dy}|p_{i,Ho}>\right) 
=    \nonumber   \hspace{4cm} \\
\left( 1 -<p_{i,Ho}|p'_{i,Dy}><p'_{i,Dy}|p_{i,Ho}> - 
\sum_{q'<F, \ne p_i'} <p_{i,Ho}|q'_{i,Dy}><q'_{i,Dy}|p_{i,Ho}>\right) 
\label{P2} 
\eea

 The physics of the two terms subtracted from 1 in eq. (\ref{P2}) is: 
The first subtracted term gives the probability, that the state $p_i'$ in Dy is occupied.
The sum in the second term takes into account the Pauli principle and prevents, that electrons 
can be moved into occupied states in Dy. The single electron states like 
$|p> = |n, \ell, j, m> $ include also 
the angular momentum projection quantum number m. This projection is for 
the description of the data irrelevant. The first subtracted term in eq. (\ref{P2}) 
gives the probability, that a specific magnetic substate $m_i'$  is occupied in $p_i$'. 
The probability, 
that all magnetic substates of $p_i$' are occupied, 
is the product of the probabilities of all substates.
One obtains the Nth. 
power of the single electron probability with $N_{p_i'} = N_{n,\ell,j; p_i'} = (2j + 1)_{p_i'}$: \\  
$ |<(n,\ell, j)_{p_i,Ho}|(n,\ell,j)_{p_i',Dy}>|^{2 N_{n,\ell,j; p_i'}}$.

\bea
P_{p_i/b} = ( 1 - |<(n,\ell,j)_{p_i, Ho}|(n,\ell,j)_{p_i',Dy}>|^{2 N_{p_i'}} - 
\nonumber \\
\sum_{(n, \ell, \ j), q_i',Dy; \ne p_i'  <F} \frac{N_{n,\ell,j} N_{n',\ell,j}}
{2j + 1}
 |<(n,\ell,j)_{p_i, Ho}|(n,\ell,j)_{q_i',Dy}>|^2  ) 
\label{P1}
\eea

With the expression: 
\be
\mbox{for n and n' with:  }  N_{n,\ell,j} = N_{n',\ell,j} = 2j +1; 
\label{P11}
\ee

 For the hole state in b' and the partially occupied state in $4f_{7/2'}$ one needs special expressions:
\be      
N_{(n,\ell,j)_b} = 2j_b \hspace{0.5cm} and \hspace{0.5cm} N_{4f7/2'} = 5.  
\label{P4}  
\ee
Here $|b>$ 
is the orbital in Ho, from which the electron is originally captured. 
The electron $p_i'$ is moved to $4f7/2$ with now 5 electrons in this orbit.  
Thus the  second (and the third) hole is in Dy in the $p_i'$ electron orbit.  
$N_{n,\ell,j}/(2j +1)$ is the averaged probability to find an electron 
in the $|p,n,\ell,j,m>$ orbital and $N_{n',\ell,j}$ the number of electron in the 
$|q_i', n',\ell,j>$ state. Primes always indicate states in Dy. 
For the probability of the second and third  hole we used the 
so called Vatai \cite{Vatai1,Vatai2} approximation with  the overlaps of 
corresponding electron wave function with the same quantum numbers in Ho and Dy equal to unity 
and neglected the exchange corrections. For the "diagonal" overlaps of 
the order of $ \approx 0.999$ this is a good approximation. 

For the probability of two different electron-particle electron-hole excitations $q_1(p_1)^{-1}$ 
and $q_2(p_2)^{-1}$ in eq. (\ref{F22}), eq. (\ref{P3}) 
one has now to introduce eq. (\ref{P2}) or finally 
eq. (\ref{P1}) as product for particle 1 and particle 2  into eq. (\ref{P3}). 

\begin{eqnarray}
P_{p_1,p_2/b} =  P_{p_1/b} P_{p_2 /b} \prod_{k=k' < F_{Dy}; \ne p_1,p_2, b} 
<k'_{Dy}|k_{Ho}>  \approx \nonumber  \hspace{5cm}\\
 ( 1 - |<(n,\ell,j)_{p_1, Ho}|(n,\ell,j)_{p_1',Dy}>|^{2 N_{p_i'}} - 
\nonumber  \hspace{4cm}\\
\sum_{(n, \ell, \ j), q_1',Dy; \ne p_1'  <F} \frac{N_{n,\ell,j,p_1,Ho} N_{n',\ell,j,q_1,Dy}}
{2j + 1}
 |<(n,\ell,j)_{p_1, Ho}|(n,\ell,j)_{q_1',Dy}>|^2  ) \nonumber \hspace{3cm} \\ 
( 1 - |<(n,\ell,j)_{p_2, Ho}|
(n,\ell,j)_{p_2',Dy}>|^{2 N_{p_2'}} -  \hspace{6cm} 
\nonumber \\                                                           
\sum_{(n, \ell, \ j)q_2',Dy; \ne p_2'<F} \frac{N_{n,\ell,j,p_2,Ho} N_{n',\ell,j,q_2,Dy}}
{2j + 1}
 |<(n,\ell,j)_{p_2, Ho}|(n,\ell,j)_{q_2',Dy}>|^2  ) \hspace{1cm}  \nonumber \\
\prod_{k=k' < F_{Dy}; \ne p_1,p_2, b} 
<k'_{Dy}|k_{Ho}>  \hspace{7cm}
\label{P6} 
\end{eqnarray}
The three hole probability normalized 
to the electron capture from the M1 $3s1/2$ state in 163 Holmium is then obtained by multiplying this 
expression into eq. (\ref{Bff}) for $P_b$. 
\vspace{1cm}

\section{Spectrum of Dy deexcitation with three-hole states.}

Tables \ref{OneHoles} and  \ref{TwoHoles} show the one and the two-hole resonance energies 
and widths  taken from results of the ECHo collaboration \cite{Ra, Ra2} and, if there not available, taken in reference 
\cite{Robertson} from the handbook of Chemistry tabulated by Weast \cite{Weast}. 
The two hole energies are determined as the one hole 
excitation energy in Dysprosium and the electron-particle and electron-hole 
energy differences in Holmium. Due to the hole in an inner shell in Dysprosium, 
which reduces the shielding of the charge of the nucleus, 
 the field looks 
for the outer electron  particle-hole excitation in Dysprosium similar 
as in Holmium with one positive charge more. This is only a rough approximation, 
since the relevant one-hole states in Dysprosium M1 3s1/2, N2 3p1/2, N1 4s1/2, 
N2 4p1/2, O1 5s1/2 and O2 5p1/2 are not from the inner most shells and thus 
the outer particle-hole states in Dysprosium do feel only approximately 
the selfconsistent field of Holmium. A better approach would be to calculate 
the electron particle-hole energies in Dysprosium selfconsistently with the hole 
in the appropriate state. This could shift the resonance energies by a few eV 
compared to the approximation used. 

The three hole resonance energy is in the approximation discussed:

\begin{eqnarray}
E_{c, (b,p_1,p_2)^{-1}} = E_{c, b^{-1}, Dy} +(E_{c,(p_1)^{-1}, Ho} - E_{c,q_1,Ho}) 
 + (E_{c,(p_2)^{-1}, Ho} - E_{c,q_2,Ho }) 
\label{ph1}
\end{eqnarray}

Where the particle-hole energy can be taken from table \ref{TwoHoles} 
starting with column four and subtracting the corresponding 
one hole energy  $E_{c,b^{-1}}$ from table \ref{OneHoles}: 

\begin{eqnarray}
E_{c, particle-hole\  energy} = (E_{c,(p_1)^{-1}, Ho} - E_{c,q_1,Ho}) 
=  \nonumber \\ E_{c,(b,p_1)^{-1}}(column\ 4, \ table \ \ref{TwoHoles}) 
- E_{c,b^{-1}}(one-hole \ energy \ of \ table\ \ref{OneHoles})
\label{ph2} 
\end{eqnarray}

The numerical calculation includes 6 Dysprosium  one-hole states $3s1/2, \ 3p1/2, \ 4s1/2,$ 
 $\ 4p1/2, \ 5s1/2 $\ and \ $ 5p1/2$ listed in table \ref{OneHoles} and 39 two-hole states tabulated in table \ref{TwoHoles}. 
The problem of choosing for the particle-hole energies 
the values of Ho is discussed before eq. (\ref{ph1}). 
The allowed 208 three hole states in this work 
are limited by energy conservation to $Q \ \ge \ 
E_C(1-hole) \ $ $+ \ E_C(particle-hole_1)\  +\  E_C(particle-hole_2)$ with the 
particle-hole excitation energies $E_C(particle-hole)$, taken as the ionization energy in Ho. 

\begin{table}
\caption{ The 6 one-hole excitation energies $E_C$ and the widths $\Gamma$ in $^{163}Dy$   
 with quantum numbers $n, \ell, j$ 
are listed according to the ECHo collaboration \cite{Blaum,Ra,Ra2,Lo2} as far as they are available. Else they are  
taken according to ref. \cite{Robertson} from Weast tabulated in the Handbook of Chemistry  \cite{Weast}, although there seem 
to be in some cases better values in the literature 
\cite{Deslattes}, \cite{Thompson}, \cite{Campbell} and \cite{Cohen}. 
 A Q value of 2.8 keV allows energetically the excitations of the  one hole states: 
M1 3s1/2, M2 3p1/2,
 N1 4s1/2, N2 4p1/2, O1 5s1/2 and O2 5p1/2.  Electrons can be captured only  from s1/2 and p1/2 
orbits in Holmium, since other 
electrons have no probability to be at the nucleus. }
\label{OneHoles}
\begin{center}

\begin{tabular}{|c|c|r|r||c|c|r|r|} \hline
$Number$ & $1.\ hole$ & $E_c [eV]$  & $\Gamma[eV] $ & $Number$& $1.\ hole$ &  $E_C$ & $\Gamma[eV]$ \\ \hline \hline
 1 & $3s1/2$ & 2040 & 13.7  &  2  &  $3p1/2$ & 1836  &  7.2   \\ \hline
3  & $4s1/2$ & 411 & 5.3 & 4 & $4p1/2$ & 333 & 8  \\ \hline
 5 & $5s1/2$ & 48 & 4.3 & 6 & $5p1/2$ & 30.8  &  3   \\ \hline \hline
\end{tabular}
\end{center}
\end{table}

\vspace{1cm}

\begin{table}
\caption{ The 39 (from number 7 to 45)  two-hole excitation energies $E_C$ and the widths $\Gamma$ in $^{163}Dy$ with quantum numbers $n, \ell, j$ 
are listed. 
We adopt here the first hole values for $E_c$ and the width $ \Gamma$  
 of the ECHo collaboration \cite{Blaum,Ra, Ra2,Lo2} and add the excitation energies of the second hole taken by Robertson \cite{Robertson}
taken from Weast tabulated in the Handbook of Chemistry  \cite{Weast}. 
The data measured by the ECHo collaboration \cite{Ra2}, \cite{Lo2} are listed in table \ref{OneHoles}.
The energies of the two hole states are given by $E_C(1 hole Dy) +
 E_C(particle-hole\  excitation \ in\  Ho)$.
 The number of one- and two-hole states listed in 
tables \ref{OneHoles} and \ref{TwoHoles} is 45. The allowed 208 three hole states included in this work 
are limited by energy conservation to $Q \ \ge \ 
E_C(1-hole) \ + \ E_C(particle-hole_1)\  +\  E_C(particle-hole_2)$. 
The particle-hole excitation energies $E_C(particle-hole)$ can be obtained 
from this table by subtracting the one-hole excitations 
given in table \ref{OneHoles}. 
\label{TwoHoles}}
\begin{center}

\begin{tabular}{|c|c|c|r|r||c|c|c|r|r|} \hline
$Number$ & $1.\ hole$ & $2.\  hole$ & $E_c [eV]$  & $\Gamma[eV] $& $Number$ &  $1.\ hole$& $2.\ hole$ 
& $E_C$& $\Gamma[eV]$ \\ \hline \hline
7 & 3s1/2 &  4s1/2 & 2472.4  & 13.7 & 27 & 4s1/2 & 4s1/2  &  841.4  &  5.4 \\ \hline
8 & 3s1/2 & 4p1/2  & 2385.3  & 13.2 & 28 & 4s1/2 & 4p1/2  &  752.5  &  5.4 \\ \hline
9 & 3s1/2 & 4p3/2  & 2350.0  & 13.2 & 29 & 4s1/2 & 4p3/2  &  717.2  &  5.4 \\ \hline
10 & 3s1/2 & 4d3/2  & 2201.8 & 13.2 & 30 & 4s1/2 & 4d3/2  &  569.0  &  5.4 \\ \hline
11 & 3s1/2 & 4d5/2  & 2201.8 & 13.2 & 31 & 4s1/2 & 4d5/2  &  569.0  &  5.4 \\ \hline
12 & 3s1/2 & 4f5/2  & 2050.4 & 13.2 & 32 & 4s1/2 & 4f5/2  &  417.6  &  5.4 \\ \hline
13 & 3s1/2 & 4f7/2  & 2047.0 & 13.2 & 33 & 4s1/2 & 4f7/2  &  414.2  &  5.4 \\ \hline
14 & 3s1/2 & 5s1/2  & 2091.1  & 13.2 & 34 & 4s1/2 & 5s1/2  &  458.3  &  5.4 \\ \hline
15 & 3s1/2 & 5p1/2  & 2072.6 & 13.2 & 35 & 4s1/2 & 5p1/2  &  439.8  &  5.4 \\ \hline
16 & 3s1/2 & 5p3/2  & 2065.9  & 13.2 & 36 & 4s1/2 & 5p3/2  &  433.1  &  5.4 \\ \hline \hline 
 17 & 3p1/2 & 4s1/2  & 2269.2  &  6   & 37 & 4p1/2 & 4p1/2  &  671.8  &  5.3 \\ \hline
18 & 3p1/2 & 4p1/2  & 2180.3  &  6   &  38 &4p1/2 & 4p3/2  &  636.5  &  5.3 \\ \hline
 19 & 3p1/2 & 4p3/2  & 2145.0  &  6   &  39 & 4p1/2 & 4d3/2  &  488.3  &  5.3 \\ \hline
 20 & 3p1/2 & 4d3/2  & 1996.8  &  6   &  40 & 4p1/2 & 4d5/2  &  488.3  &  5.3 \\ \hline
 21 & 3p1/2 & 4d5/2  & 1996.8  &  6   &  41 & 4p1/2 & 4f5/2  &  336.9  &  5.3 \\ \hline
 22 & 3p1/2 & 4f5/2  & 1845.4  &  6   &  42 & 4p1/2 & 4f7/2  &  328.3  &  5.3 \\ \hline
 23 & 3p1/2 & 4f7/2  & 1842.0  &  6   &  43 & 4p1/2 & 5s1/2  &  377.6  &  5.3 \\ \hline
 24 & 3p1/2 & 5s1/2  & 1886.1  &  6   &  44 & 4p1/2 & 5p1/2  &  359.1  &  5.3 \\ \hline
 25 & 3p1/2 & 5p1/2  & 1887.6  &  6   &  45 & p1/2 & 5p3/2  &  352.4  &  5.3 \\ \hline
 26 & 3p1/2 & 5p3/2  & 1860.9  &  6   & ---& ---& --- & ---& \\ \hline \hline
\end{tabular}
\end{center}
\end{table}

\begin{figure}[tp]
\begin{center}
\begin{minipage}[tl]{18 cm}
\epsfig{file=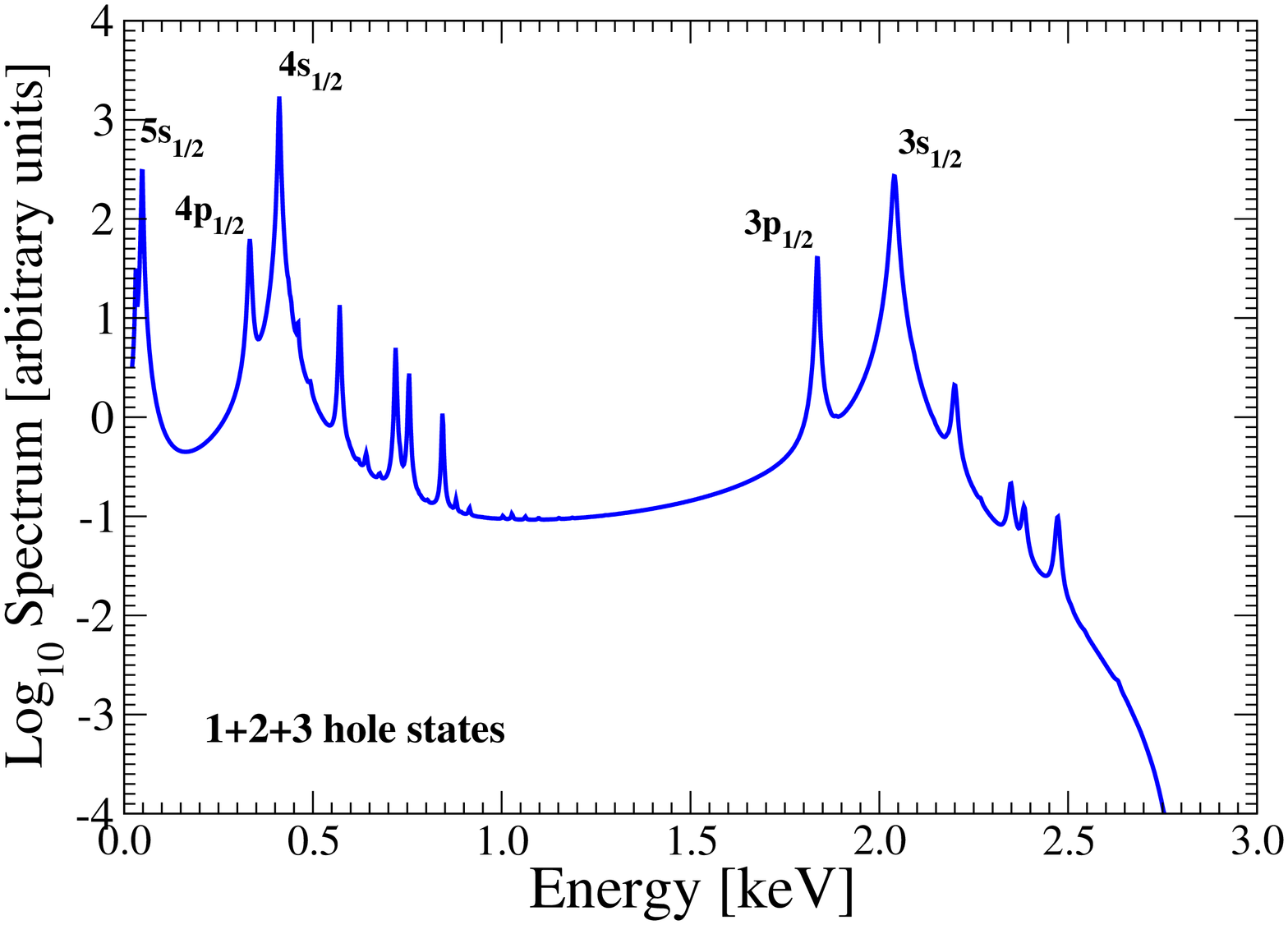,scale=0.7,angle=-90}
\end{minipage}
\begin{minipage}[t]{16.5 cm}
\caption{(Color online) $Logarithmic_{10}$ three-hole Bolometer Spectrum (\ref{decay}) 
with (\ref{Bff}) and  (\ref{P6}) for the sum of the  one-, 
the two- and the three-hole probabilities calculated in this work  
with the resonance energies and widths of tables \ref{OneHoles} and  \ref{TwoHoles} with the assumed Q-value Q = 2.8 keV for
the bolometer energy between 0.0 and 2.8 keV. The $logarithmic_{10}$  coordinates of the 
ordinate has to be read as $10^{ordinate}$, e. g. the ordinate value -2 corresponds to $10^{-2}$. 
The theoretical spectra are normalized to the experimental $4s_{1/2}$ hole peak at 0.411 keV
(see figure \ref{Three-Fig-4}).
\label{Three-Fig-1}}
\end{minipage}
\end{center}
\end{figure}

\begin{figure}[tp]
\begin{center}
\begin{minipage}[tl]{18 cm}
\epsfig{file=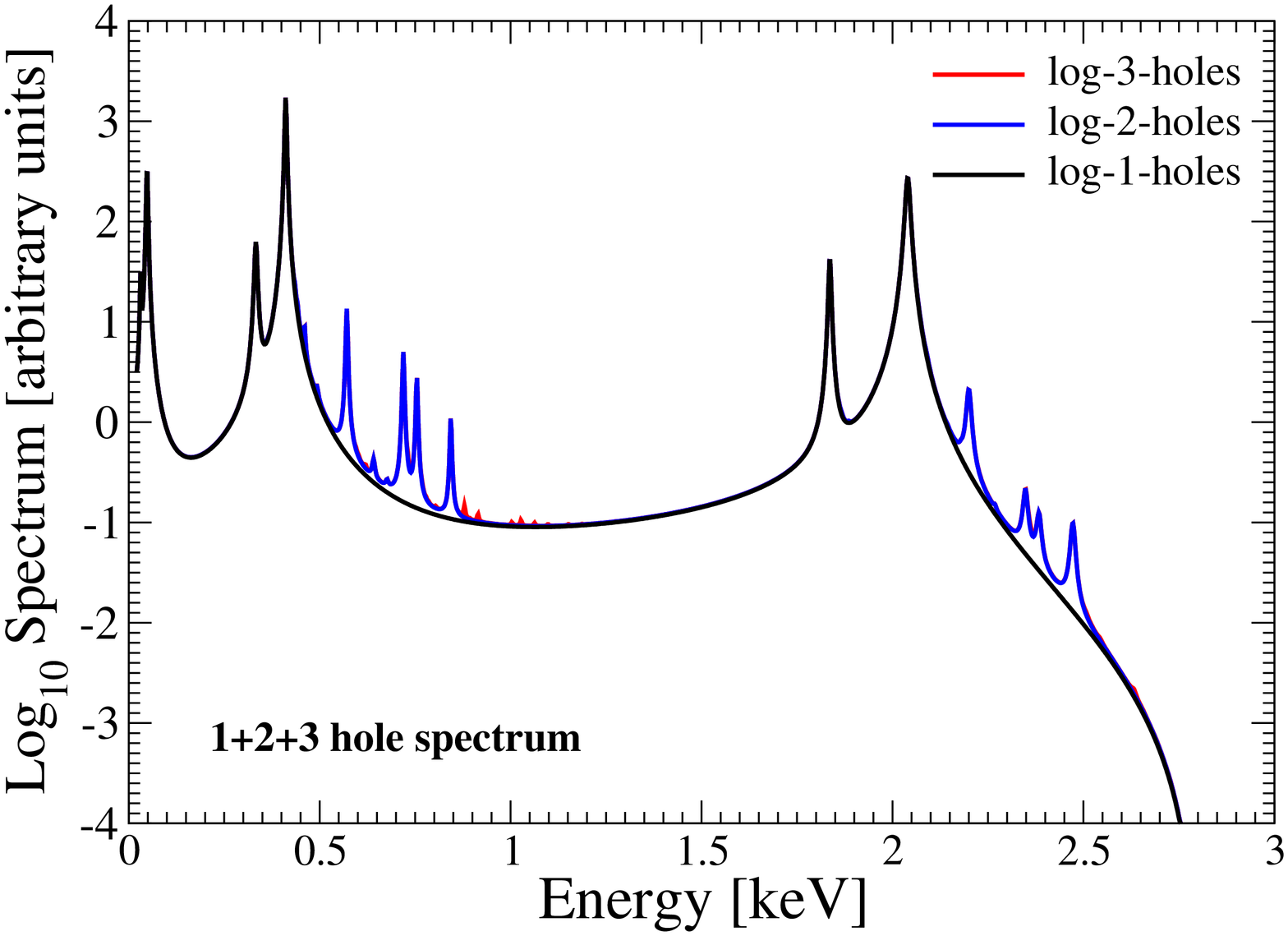,scale=0.7,angle=-90}
\end{minipage}
\begin{minipage}[t]{16.5 cm}
\caption{(Color online) $Logarithmic_{10}$ plot of the bolometer spectrum (\ref{decay})  including the one-, 
the one- plus two- and the one- plus two- plus three-hole states. The Q value 
is assumed to be Q = 2.8 keV according to an ECHo measurement \cite{Ra}. The resonance energies and widths 
are listed in tables \ref{OneHoles} and  \ref{TwoHoles}. The three-hole states can hardly be seen.
\label{Three-Fig-2}}
\end{minipage}
\end{center}
\end{figure}

Figure \ref{Three-Fig-1} shows the sum of the one-, two- and three-hole 
theoretical bolometer spectrum after electron capture in Holmium for the deexcitation 
of Dysprosium. The electron wave functions are calculated in 
a Dirac-Hartree-Fock approach 
\cite{Grant, Desclaux, Aukudinov}. In Dysprosium the one-hole states are included 
in the selfconsistent determination of the electron wave functions. 
A detailed description of our approach is given in ref. \cite{Fae4}.  Figure  \ref{Three-Fig-2} 
compares the sum of the theoretical one- plus two- plus three-hole deexcitation spectrum of Dysprosium with the results including 
only one-hole states  and the sum of the one- and the two-hole states. The three-hole contributions can practically not be seen even in this $logarithmic_{10}$ plot.

\begin{figure}[tp]
\begin{center}
\begin{minipage}[tl]{18 cm}
\epsfig{file=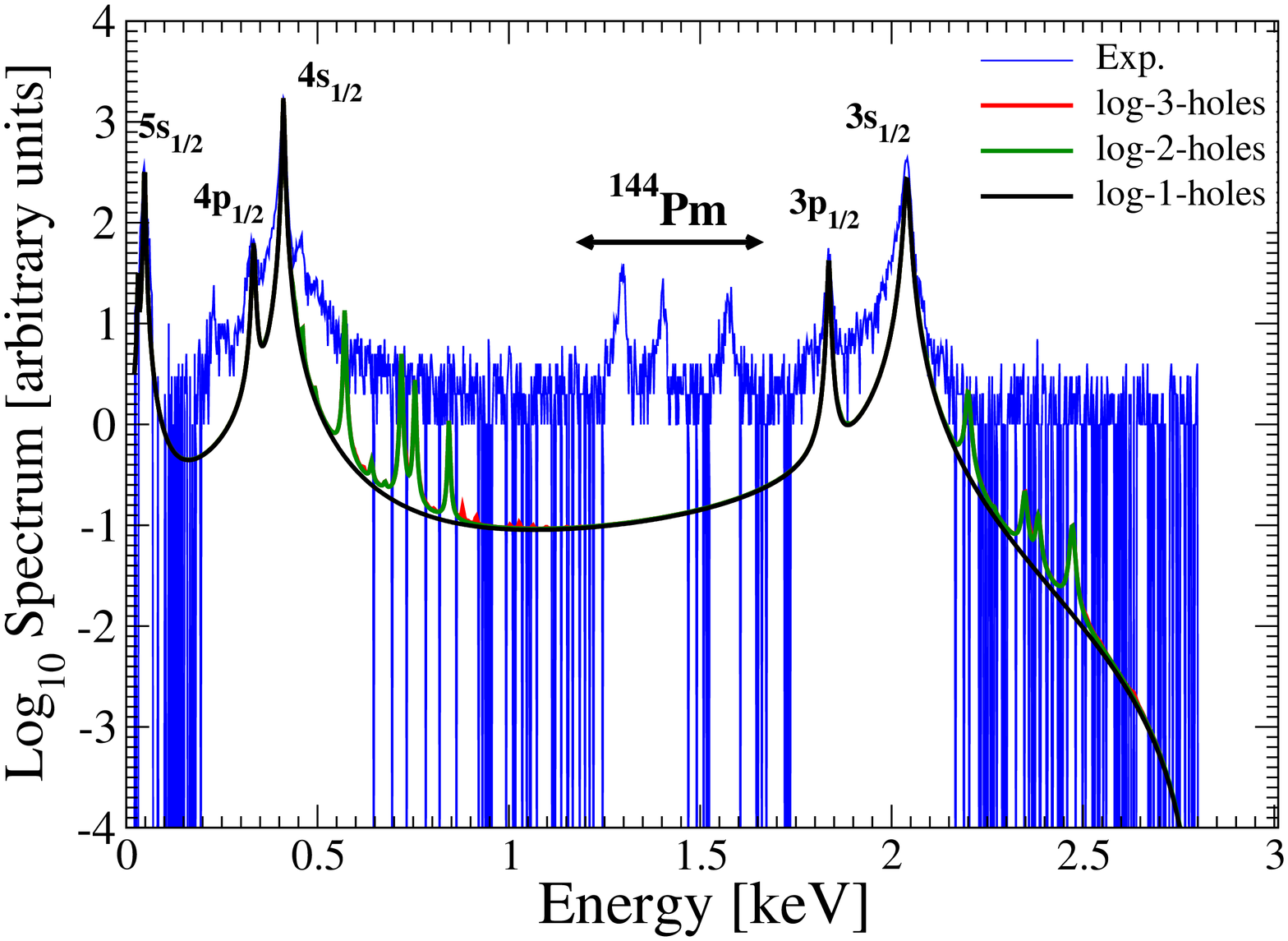,scale=0.7,angle=-90}
\end{minipage}
\begin{minipage}[t]{16.5 cm}
\caption{(Color online) Comparison of the $logarithmic_{10}$ spectrum of the one-, 
the one- plus two and the one- plus two- plus    three-hole states 
with the data of the ECHo collaboration 
\cite{Blaum, Ra, Ra2, Lo2}. The experimental data are binned in 2 eV (counts/2eV). 
Some bins have no counts and thus the 
logarithmic value is $-\infty$. The effect of the three-hole states does show hardly up even in this logarithmic figure. 
\label{Three-Fig-3}}
\end{minipage}
\end{center}
\end{figure}

\begin{figure}[tp]
\begin{center}
\begin{minipage}[tl]{18 cm}
\epsfig{file=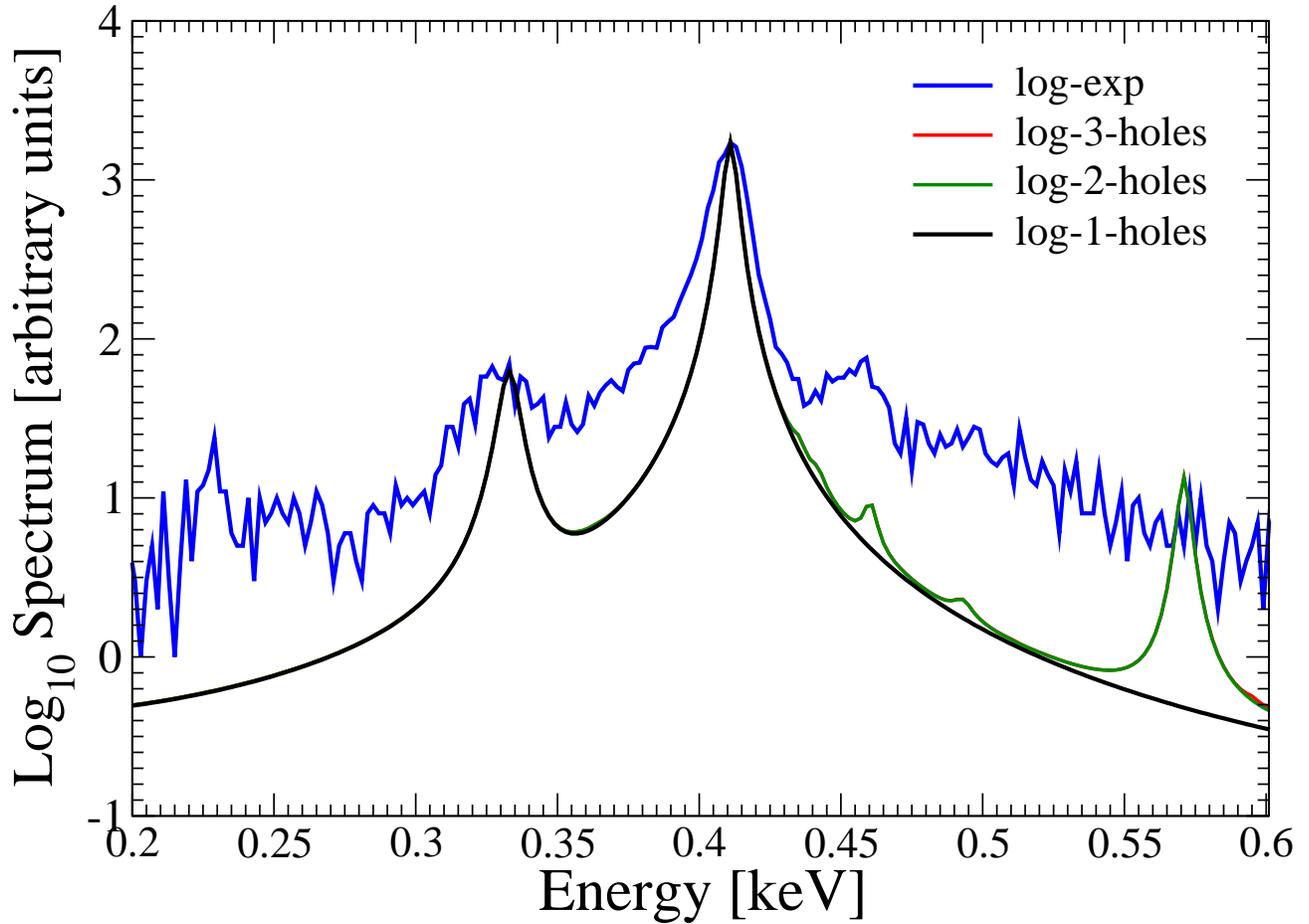,scale=0.7,angle=-90}
\end{minipage}
\begin{minipage}[t]{16.5 cm}
\caption{(Color online) Comparison of the $logarithmic_{10}$ bolometer spectrum around the 
$N1 \ = \ (4s1/2)^{-1}\  0.411 \ keV$ (relative probability to the 3s1/2 state  24.4 \%) and $N2 \ = \ (4p1/2)^{-1} \  0.333\  keV$ (relative probability 1.22 \% )  
resonances with the ECHo data 
\cite{Blaum, Ra, Ra2, Lo2}. At 0.569 keV one sees the degenerate two-hole states $N1 \ = \ (4s1/2)^{-1}$, $N4 \ = \ (4d3/2)^{-1}$ with the relative probability of 0.088 \% and 
$N1 \ = \ (4s1/2)^{-1}$, $N5 \ = \ (4d5/2)^{-1}$ with the relative probability
 of 0.125 \% (total 0.213 \%). The  resonance energies and the widths are taken from the ECHo collaboration \cite {Ra2,Lo2}. 
The theory does not include the 
finite resolution of the measurement, which is about 10 eV FWHM. 
The sum of the one- plus two- and the sum of the one- plus two- plus three-hole spectra look even in a logarithmic scale identical. 
\label{Three-Fig-4}}
\end{minipage}
\end{center}
\end{figure}

Figure \ref{Three-Fig-3} compares in a $logarithmic_{10}$  plot the deexcitation spectrum 
of the one-hole states, 
the one- plus two-hole and the one- plus two- plus three-hole excitations with 
the ECHo data \cite{Ra2,Lo2}. The experimental spectrum in binned into 2 eV. Since 
some bins contain no counts the logarithmic value is $-\infty$.  
For a more relevant logarithmic plot one would need more counts. The three 
experimental peaks between 1.2 and 1.6 keV originate from a $^{144}Pm$ contamination. 
Figure \ref{Three-Fig-4} shows in detail a comparison of the one-, the two- and 
three-hole spectrum with the ECHo data \cite{Ra2,Lo2} around the area of the 
N2 4p1/2, 0.333 keV and the N1 4s1/2, 411 keV hole states. 

\section{Conclusions}

Here for the first time the effect of three hole excitations on the deexcitation 
spectrum of $^{163}Dy$  after electron capture in $^{163}Ho$ has been studied. 
The spectrum is not  affected visibly by the three-hole excitations of   electron hole and two 
electron particle-hole states. 
The two electron particle-hole excitations plus the energy of a one-hole 
state (three-hole states) add up to higher energies as the 
 one- and two-hole states. 
A one-, two- or three-hole state close to the Q value could complicate 
the determination of the neutrino mass. If only one resonance state is dominant near the Q value and if the line profile is Lorentzian (or of an other known analytical form with two parameters), one needs to fit 
simultaneously  four parameters of the theory to the data: (1) neutrino mass, 
(2) the energy difference of the dominant state to the Q value, (3) the width of the resonance and (4) the strength. Before fitting the four parameters of the theoretical spectrum to the data, one has to fold the experimental spectral function of the detector into the theoretical results. The situation is more 
complicated, if several one- or two- or three-hole resonances contribute to the shape of the spectrum near the Q value.  With the recent measurement \cite{Ra2,Lo2} 
of the ECHo collaboration
(\ref{QECHo}) $Q = 2.80\ \pm \  0.08\ [keV]$, 
it has been shown by Faessler and Simkovic \cite{Fae4}, that the highest one hole state 
is dominant at the Q value. The high energy two-hole states do not influence the behavior in the region, which determines the neutrino mass \cite{Fae4}. 
Although the three-hole excitations can come closer to the Q value, their strength is so weak, that they are not dangerous for the determination of the neutrino mass.  
 
\vspace{1cm}

Acknowledgment: We want to thank other members of the ECHo collaboration for discussions.

\end{document}